\documentclass[conference]{IEEEtran}
\IEEEoverridecommandlockouts
\usepackage{cite}
\usepackage{amsmath,amssymb,amsfonts}
\usepackage{algorithmic}
\usepackage{graphicx}
\usepackage{textcomp}
\usepackage{xcolor}
\def\BibTeX{{\rm B\kern-.05em{\sc i\kern-.025em b}\kern-.08em
    T\kern-.1667em\lower.7ex\hbox{E}\kern-.125emX}}

\usepackage{ifthen}
\usepackage{hyperref}

\newboolean{forreview}
\setboolean{forreview}{false}
\ifthenelse{\boolean{forreview}}
{\newcommand{\ifReview}[2]{{#1}}}
{\newcommand{\ifReview}[2]{{#2}}}

\makeatletter
\newcommand{\linebreakand}{%
  \end{@IEEEauthorhalign}
  \hfill\mbox{}\par
  \mbox{}\hfill\begin{@IEEEauthorhalign}
}
\makeatother

\begin{document}
\title{Horizontal Scaling of Transaction-Creating Machines

}
\ifReview{}{
\author{
\IEEEauthorblockN{1\textsuperscript{st} Ole Delzer}
\IEEEauthorblockA{\textit{Technische Universitaet Berlin} \\
ole.delzer@tu-berlin.de
}
\and
\IEEEauthorblockN{2\textsuperscript{nd} Ingo Weber}
\IEEEauthorblockA{\textit{Technical University of Munich} \\
\textit{\& Fraunhofer Gesellschaft} \\
ingo.weber@tum.de
}
\and
\IEEEauthorblockN{3\textsuperscript{rd} Richard Hobeck}
\IEEEauthorblockA{\textit{Technische Universitaet Berlin} \\
richard.hobeck@tu-berlin.de}
\linebreakand
\IEEEauthorblockN{4\textsuperscript{th} Stefan Schulte}
\IEEEauthorblockA{\textit{Hamburg University of Technology} \\
stefan.schulte@tuhh.de}

}
}

\maketitle

\begin{abstract}
Blockchain technology has become one of the most popular trends in IT over the last few years. Its increasing popularity and the discovery of ever more use cases raises the question of how to improve scalability. While researchers are exploring ways to scale the on-chain processing of transactions, the scalability of the off-chain creation of transactions has not been investigated yet.
This is relevant for organizations wishing to send a high volume of transactions in a short time frame, or continuously, e.g., manufacturers of high-volume products.
Especially for blockchain implementations such as Ethereum, which require transactions to include so-called nonces (essentially a sequence number), horizontally scaling transaction creation is non-trivial. 
In this paper, we propose four different approaches for horizontal scaling of transaction creation in Ethereum. 
Our experimental evaluation examines the performance of the different approaches in terms of scalability and latency and finds two of the four proposed approaches feasible to scale transaction creation horizontally. 
\end{abstract}


\section{Introduction}
\label{sec:intro}


Since the introduction of expressive smart contracts, blockchain has been explored as a foundation for decentralized applications (dapps) and multi-party business processes in a lot of industries~\cite{wood2014ethereum,2019-Bratanova-ACS}.
Early blockchain platforms, like Bitcoin and Ethereum, offered maximal transaction throughput rates between 3 and 15 transactions per second (tps) initially~\cite{2017-Weber-SRDS}. Limited transaction throughput is widely discussed as a factor impeding the scalability of blockchain applications~\cite{khan2021systematic} -- for recent advances see \autoref{sec:relw}.
This very limited throughput scalability motivated numerous alternative proposals of consensus algorithms and blockchain platforms, including
Ripple with 1\,500~tps~\cite{bach18} and
the RedBelly Blockchain with 30k~tps~\cite{Crain:2021:RBBC}. 
As per the ``Blockchain Trilemma'', typically high throughput scalability comes as a tradeoff against the properties security and decentralization~\cite{Trilemma}.
Still, blockchain platforms like the above can be used in business scenarios, such that throughput scalability to thousands of tps can be taken as readily available.

In some blockchain application scenarios, organizations may need to send a high volume of transactions in a short time frame, or continuously. 
This includes among others manufacturers of high-volume products, e.g., in application scenarios like traceability of food or pharmaceutical products.
For instance, the world's largest beverage companies bottle hundreds of millions of products per day, and the world's largest fast food companies produce similar amounts of meat products per day.
To register each product with an individual identifier (ID) on a blockchain,
per 100 million transactions (tx) that need to be processed per day, a throughput of approx.\ 1158 tps results (assuming 1 tx per ID).
In such applications, a single actor \ifReview{}{(the producer) }needs to issue transactions at such a throughput rate, to enable use cases like food traceability at scale.
Such use cases typically span supply chains, which involve many actors in addition to manufacturers of consumer foods, and arguments for their decentralized implementation using blockchain technologies apply in many (though not all) cases.

However, conventional approaches to creating transactions do not scale easily, within a single machine, to such throughput rates.\footnote{For an individual use case, alternative architectures can be designed; however, we here address the general class of problems where \textit{high throughput in transaction creation} is required.
Layer 2 technologies are often no viable solution, if the goal is to create persistent records on a blockchain.}
Thus there is a need to scale \ifReview{tx creation}{transaction-creating machines} horizontally.
This is non-trivial, though, since blockchains typically require a unique identifier for each transaction; the unique identifier is required to prevent replay attacks.
For instance, in Ethereum, this is achieved with the so-called \textit{nonce}, essentially a sequence number for transactions created by a given sender account.
The combination of sender account and nonce is unique.

Horizontal scaling of a function that relies on a shared variable is in general not a new topic, but the specific setting for blockchain transaction creation allows for different solution approaches than other settings. 
To the best of our knowledge, this is the first paper formulating and addressing this problem.

Our contribution is as follows.
In this paper we propose four alternative approaches for achieving such scalability, 
in part making use of the specifics of the environment like employing smart contracts.
We implemented the approaches, made the code available, and conducted experiments to study and contrast the properties of the four approaches, particularly in terms of scalability, latency, and fairness.
Summarizing the evaluation results, we find that one approach only scales sub-linearly, and another one is typically less preferable than a third. In our experiments, the remaining two approaches scaled well, offered high throughput rates, and -- with suitable parameter setting -- achieved good fairness regarding the distribution of transaction inclusion latency.

\ifReview{In the following, after}{The remainder of the paper is structured as follows. After} discussing related work in \autoref{sec:relw}, we present the four alternative approaches in \autoref{sec:section3}. The evaluation is presented in \autoref{sec:analysis}, and the results are discussed in \autoref{sec:discussion} before \autoref{sec:conclusion} concludes.


\section{Related Work} 
\label{sec:relw}
Blockchain technology itself is can still be considered novel to some extent and its broader, widespread use is just starting. Hence, research on scaling in blockchain technologies has only started some years ago. In 2016, Croman et al.\ were among the first researchers to scientifically explore ways for scaling blockchains~\cite{croman2016}. They also explained why earlier methods for scaling -- increasing the block size so that more transactions can be included in a single block while decreasing the inter-block time -- are limited. That is why other, more drastic changes are necessary to enable scalability to industrial use cases. Another early paper by Vukoli{\'{c}}~\cite{Vukolic2016} discussed different proposals for blockchain scalability, including the option to rely on alternatives to Proof-of-Work (PoW).

Alongside Bitcoin, other cryptocurrencies such as Ethereum also experienced increasing interest. Ethereum in particular popularized 
Smart Contracts, which enabled a much broader application of blockchain technology than ``just'' value transfer, but also increased the complexity of transactions. Of course, this increased complexity also needs to be accounted for in possible approaches for scaling blockchain technology. For example, Dickerson et al.~\cite{Dickerson2017} propose that miners speculatively process transactions (and thereby execute Smart Contracts) concurrently. If conflicts arise, the affected contracts are rolled back and serially re-executed. The resulting execution schedule is populated alongside the mined block so that validators can execute the Smart Contracts in parallel deterministically.


Early blockchain protocols achieved only a limited number of tps. Regarding concrete blockchain platforms and throughput numbers, for instance, Bitcoin achieves ca. 7~tps.
Ethereum achieved 15 tps in 2018~\cite{bach18} and could, due to increased block ``size'' and frequency at the time of writing, achieve a theoretical throughput of approx.\ 120 tps\footnote{With a block gas limit of about 30 Mio.\ gas, a block approx.\ every 12 seconds, and a minimum of 21k gas per transaction (see \url{https://etherscan.io/chart/gaslimit}, accessed 2022-12-16), approx.\ 118 tps would be possible. Actual throughput depends on transaction complexity and demand, and is typically lower.}. More recent protocols achieve higher numbers, e.g., Ripple aims at 1\,500~tps, and EOS aims at millions of tps~\cite{bach18}. The RedBelly Blockchain achieved 30k~tps in a globally distributed network~\cite{Crain:2021:RBBC}, with the authors stating that the bottleneck in this experiment was the load generation.
Importantly, the possible tps are primarily affected by the applied consensus mechanism. The usage of PoW naturally leads to a low tps, while \ifReview{}{(Delegated)} Proof-of-Stake \ifReview{(PoS)}{} enables increasing the tps to higher numbers. For instance, EOS implements Delegated \ifReview{PoS}{Proof-of-Stake}, while the upcoming Ethereum~2.0 applies \ifReview{PoS}{Proof-of-Stake}, and aims at 200k to 300k~tps. 

The discussed approaches mainly aim at increasing the transaction throughput of blockchains and therefore only consider the on-chain transaction processing by miners and validators. For the client-side of dapps, i.e., for the creation of transactions, these approaches are not applicable. In fact, so far there does not seem to be any scientific research regarding the scalability of \ifReview{tx}{transaction} creation. 

Furthermore, the area of blockchain benchmarking is also relevant to the work at hand. Benchmarking of blockchains with regard to scalability and general performance has been an important research topic in recent years~\cite{wang19}. For instance, Gervais et al.~\cite{gervais16} present a simulation framework for analyzing security and performance constraints of PoW blockchains. Dinh et al.~\cite{dinh17} implement BLOCKBENCH, which is a framework used to analyze private blockchains. For this, respective workloads are defined. Another benchmarking framework for blockchains is BCTMark~\cite{saingre20}. In contrast to BLOCKBENCH, BCTMark also allows to target non-private blockchains. 
Finally, Hyperledger Caliper\footnote{\url{https://hyperledger.github.io/caliper/}, accessed 2022-06-22} is used to benchmark the performance of Hyperledger-based blockchains. To the best of our knowledge, none of the discussed solutions takes into account the off-chain creation of transactions. Therefore, the work at hand could be used in order to extend the existing frameworks and simulators.

\section{Approaches}
\label{sec:section3}

In this section, we describe our four different approaches to horizontally scale transaction creation.
Vertical scaling a single machine, i.e., using a faster/stronger machine, is limited by the maximal speed / power of that single machine.
Hence we focus on horizontal scaling, i.e., using more (or fewer) machines to provide a given function or service.

All four approaches are designed to accept requests from applications, which may be distributed over the available transaction (TX) creating machines, e.g., by a load balancer component (not discussed below).
These TX creating machines operate as part of the off-chain backend, which interacts with on-chain backend components like Smart Contracts.
Summarizing the four approaches, Approach 1 includes a single middleware component which takes incomplete TXs from the TX creating machines, adds the nonce, signs the TXs, and sends them to the blockchain.
In Approach 2, each TX creating machine has its own blockchain address, and can issue TXs independently; authorization of these accounts to act on behalf of e.g., some company is handled by a Smart Contract.
In Approaches 3 and 4, the TX creating machines request nonces from a singleton Nonce Manager component, either for each TX individually (Approach 3) or for contingents of TXs (Approach 4).
The approaches are presented in detail below.

\subsection{Approach 1} \label{sec:approach1}
The first approach is arguably the most straight-forward one of the four.
As depicted in Fig.~\ref{fig:approach1}, the on-chain backend consists of a Smart Contract and a user account. The Smart Contract implements the business logic, e.g., a registry of unique product IDs. The user account represents an organization, e.g., the producer of high-volume goods for which individual IDs are needed, and is used to create transactions which call the Smart Contract. 

\begin{figure*}[t]
    \centering
    \includegraphics[width=.68\textwidth]{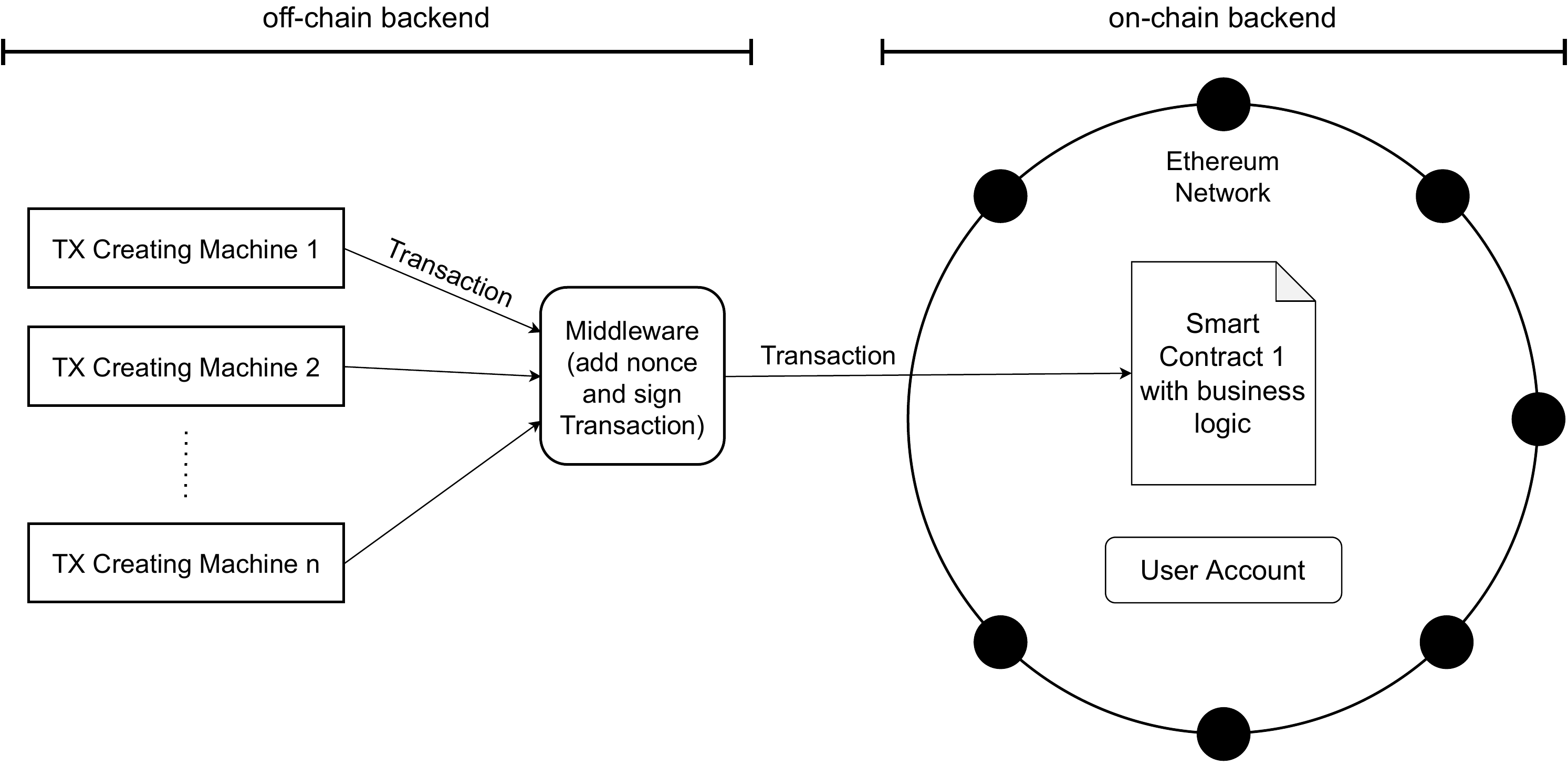}
    \caption{Approach 1: Horizontal Scaling by externalizing the nonce setting and signing the transaction in a separate, dedicated \emph{middleware}. 
    }
    \label{fig:approach1}
\end{figure*}

The off-chain backend includes the TX creating machines as per above, and a \emph{middleware} component between the TX creating machines and the blockchain system. The middleware receives transaction objects from the TX creating machines. These transaction objects are transmitted to the middleware without the nonce or signature. The middleware then adds the current nonce, signs the transaction on behalf of Ethereum user account, and forwards it to the Ethereum network. Note that the TX can only be signed once the nonce has been set. This design allows us to keep track of the nonce locally at the middleware. We only fetch the current nonce of our user account from the Ethereum Network once at startup and store it as a local variable. Then we simply increment it by one every time the middleware finalizes a new transaction. Since the nonce is irrelevant when the transaction object is first created, we can horizontally scale the transaction-creating machines without concern for the nonce.

The downside of this approach is that, while transactions can be created concurrently from multiple machines, the middleware is a singleton: setting the nonce and signing the transactions are done on a single machine. There will be a limit regarding the number of transactions the middleware can process per second, which will conceivably pose a bottleneck.

\subsection{Approach 2} \label{sec:approach2}

Fig.~\ref{fig:approach2} depicts the second approach to realize horizontal scaling of transaction creation. 
In this design, we circumvent the nonce problem, by equipping each transaction-creating machine with its own dedicated user account to sign transactions. This way, all machines can keep track of their individual nonce in a local variable. 
They can create transactions completely independent from each other, which allows for easy scaling without concerns about synchronization in the off-chain backend. 

\begin{figure*}[t]
    \centering
    \includegraphics[width=.55\textwidth]{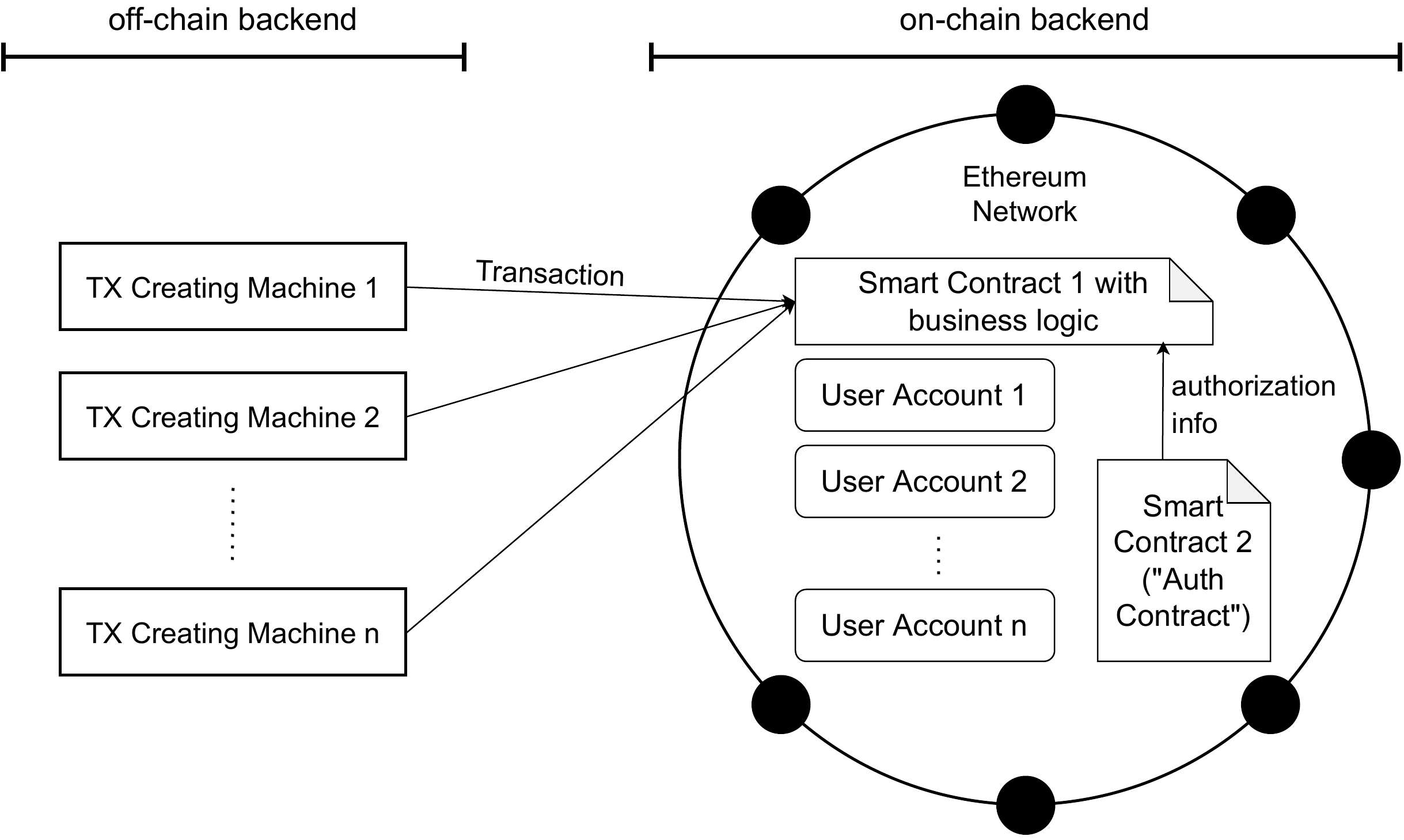}
    \caption{Approach 2: Horizontal scaling by allocating an individual account to each transaction-creating machine}
    \label{fig:approach2}
\end{figure*}

However, using multiple accounts introduces a new problem concerning \emph{authorization}: the Smart Contract containing the business logic has to be able to decide which invocations to accept (cf.\ the Embedded Permission pattern~\cite{2018-Xu-EuroPLoP}). In all other approaches, we use a single designated user account, so the Smart Contract simply checks if a transaction originates from this account. 
In contrast, here we want to dynamically scale our transaction-creating machines according to the current workload, hence the number of accounts that should be permitted to send transactions to our Smart Contract cannot be static. 
It must be possible to increase or decrease the number of authorized accounts as needed, and the Smart Contract needs to be able to verify both authorization of the accounts.
(Authentication is given by the blockchain / TX signature.)

We achieve this by introducing a second Smart Contract, which we refer to as \emph{Auth Contract} in the following. The Auth Contract maintains a list of the addresses of all currently authorized user accounts. It offers a function to check if a specific address is contained in this list, i.e., belongs to an authorized user account. The list can be updated dynamically with a dedicated \emph{master account}. Upon being invoked, our first Smart Contract containing the business logic calls the Auth Contract to check whether the transaction's sender account is authorized. Only then will the business logic of our first Smart Contract be executed; otherwise, an error is logged and no other state change takes place.

The disadvantages of this approach are the extra complexity introduced by using a variable number of distinct user accounts, and slightly higher gas usage (for deploying the Auth Contract, updating the variables, and conducting the checks). The list containing the authorized accounts' addresses has to be updated whenever the number of transaction-creating machines changes, and the key pairs for all accounts have to be managed well and kept private. 
Given that we design solutions for handling thousands of tps, we presume that the additional gas usage will be negligible, and do not consider it strongly in the remainder.

\subsection{Approach 3} \label{sec:approach3}

\begin{figure*}[t]
    \centering
    \includegraphics[width=.7\textwidth]{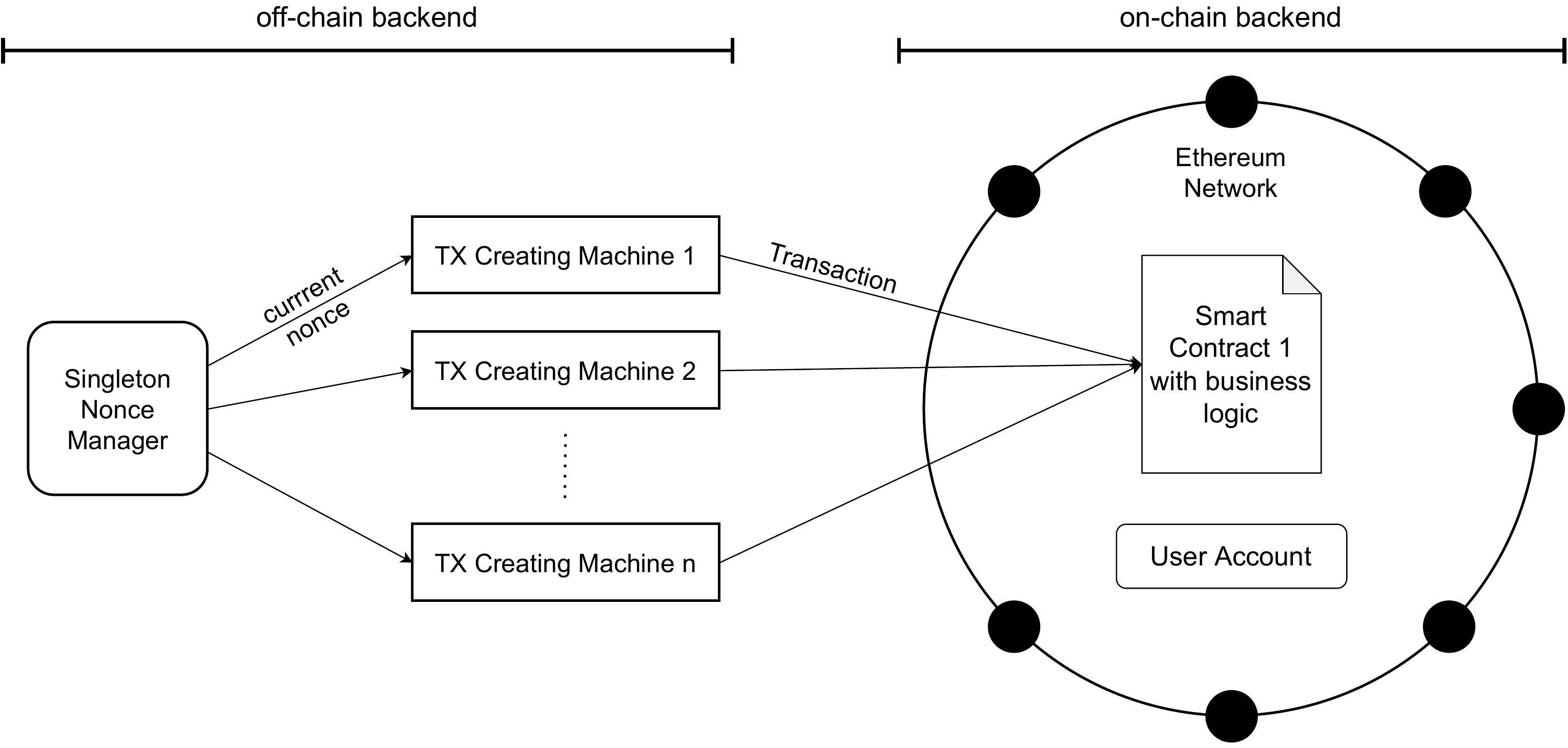}
    \caption{Approach 3: Horizontal scaling by outsourcing the nonce management to an additional singleton component called \emph{Nonce Manager}}
    \label{fig:approach3}
\end{figure*}

For Approach 3, to horizontally scale our transaction-creating machines, we outsource the nonce management to a new singleton component called \emph{Nonce Manager} (see Fig.~\ref{fig:approach3}). As a singleton, there is only one instance that is shared by all TX creating machines. So instead of maintaining the current nonce's value locally at the TX creation machines, it is only stored at the Nonce Manager. Every time the TX creating machines create a new transaction, they request a nonce value from the Nonce Manager. The Nonce Manager then responds with the value of its nonce counter and increments it by one afterwards. The transaction-creating machine sets the nonce to the received value for the new transaction, which is then signed and sent to the Ethereum network.

The disadvantage of this approach is that it entails a service invocation every time a transaction is created. This will presumably increase the time it takes to create a transaction, thereby decreasing the transaction throughput. The impact will likely depend on the network latency between a transaction-creating machine and the Nonce Manager and the load of the latter: the Nonce Manager is shared by all transaction-creating machines, and naturally subject to limited bandwidth and computing power. The Nonce Manager also poses a single point of failure. However, the Nonce Manager only implements the functionality of a data store containing a single key-value pair, so it should be feasible to achieve high performance and resiliency.

\subsection{Approach 4} \label{sec:approach4}
As we can see in Fig.~\ref{fig:approach4}, Approach 4 is quite similar to Approach 3: we also externalize nonce management to a dedicated singleton Nonce Manager. But instead of having the TX creating machines request the nonce from the Nonce Manager for every new transaction individually, we use the Nonce Manager to allocate \emph{nonce contingents} to the machines.

\begin{figure*}[t]
    \centering
    \includegraphics[width=.7\textwidth]{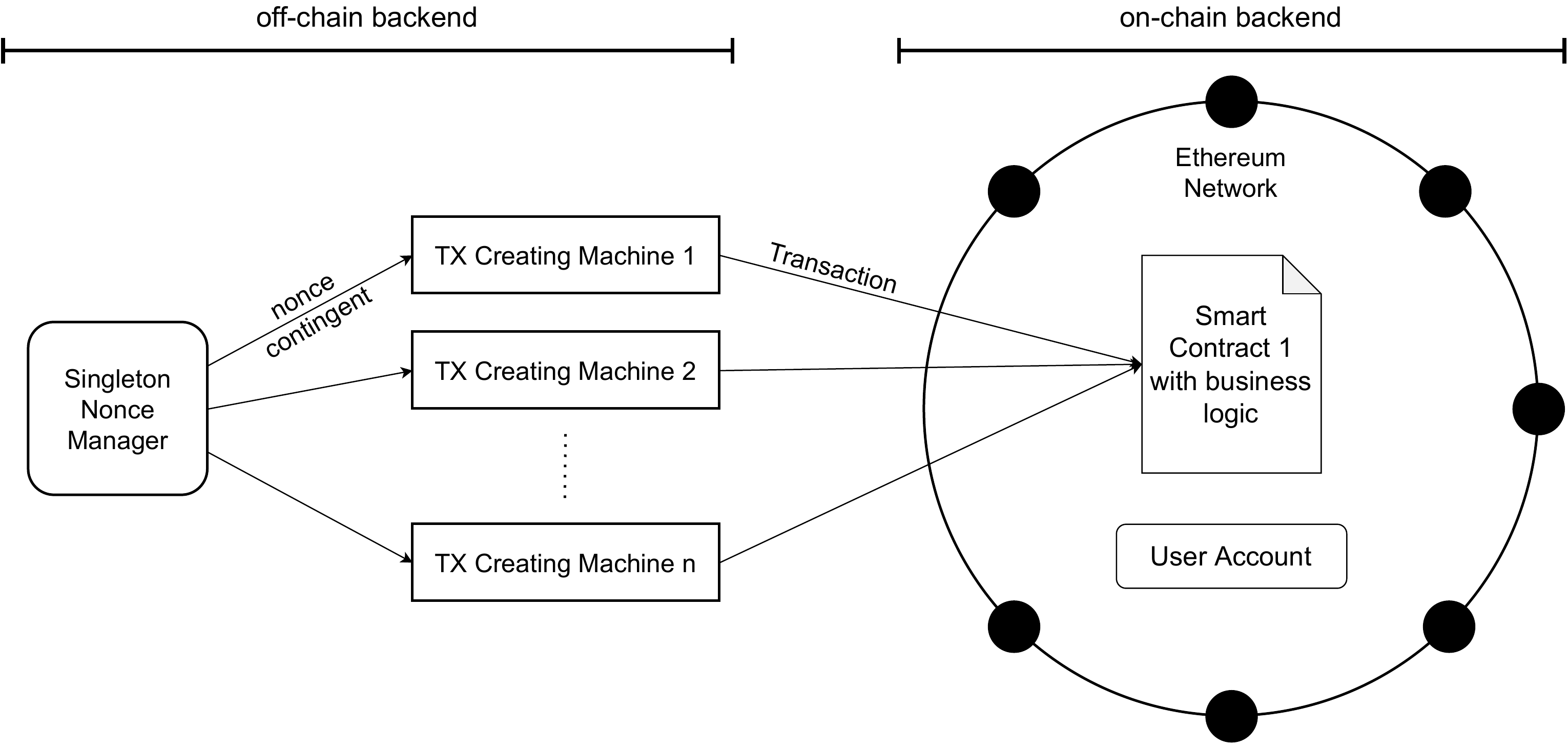}
    \caption{Approach 4: Horizontal scaling by assigning \emph{nonce contingents} to the transaction-creating machines through the \emph{Nonce Manager}}
    \label{fig:approach4}
\end{figure*}

For example, say we use a contingent size of $c=100$ and the current nonce value stored in the Nonce Manager is $1\,500$. If a particular transaction-creating machine, \emph{A}, requests the next nonce contingent from the Nonce Manager, the latter responds with the contingent of nonces $1\,500$ to $1\,599$. The Nonce Manager then increments the current nonce by the contingent size of $c=100$, so the new value is $1\,600$. \emph{A} can now start to create transactions with nonces from $1\,500$ to $1\,599$. Afterwards, \emph{A} requests another nonce contingent from the Nonce Manager again and the cycle repeats.

This will lead to situations where, for example, another transaction-creating machine, \emph{B}, creates transactions with  nonce $1\,600$ and above while \emph{A} has not yet used all nonces between $1\,500$ and $1\,599$, i.e., there are nonces lower than $1\,600$ yet unused. For transaction creation itself, this is not a problem. But once it is sent to the Ethereum network, a transaction with a nonce that is higher than a yet-unused nonce has to wait in the miner's transaction pool -- until all lower nonce values are used in a mined or pending transaction. Hence, larger contingent sizes will increase the expected waiting time (latency) of the transactions and also lead to a higher variance in the distribution of waiting times. The advantage compared to Approach 3 is that the frequency and volume of requests to the Nonce Manager is much lower when requesting whole nonce contingents instead of single nonces, which allows our Nonce Manager to serve more transaction-creating machines.
\section{Evaluation}
\label{sec:analysis}

To evaluate each of the four approaches we conducted experiments. In the experimental setup, up to three transaction-creating machines were used on individual VMs. The private Ethereum network, consisting of a single Geth node, was operated on another, separate VM. To allow high transaction throughput, the node was operated with the consensus mechanism Proof of Authority (PoA), at 400 M gas per block, and an inter-block time of 5 seconds, resulting theoretically in up to ~3800 tps. This proved sufficient for all the experiments we conducted, i.e., the blockchain was not the bottleneck. There also was a separate VM for the Middleware in Approach 1 and for the Nonce Manager in Approaches 3 and 4, respectively. We used Microsoft Azure as our cloud computing platform. All VMs were of the size \emph{Standard\_A1\_v2}, which comprises a single vCore\footnote{Intel Xeon Platinum 8272CL CPU @ $2.60$ GHz}, $2$ GiB of RAM, a download bandwidth of $1\,500$ Mbit/s and an upload bandwidth of $250$ Mbit/s.

We configured the mining node to mine a new block every $5$ seconds. For Approach 1, the total number of transaction-creating machines has an impact on their performance, so we tested Approach 1 with three different configurations: with only one machine ($m=1$), with two ($m=2$), and with three ($m=3$) machines. Approach 4 was also tested with different configurations, at first with a nonce contingent size of $c=100$, then $c=1\,000$, and lastly, $c=10\,000$. Simulations for Approaches 2 and 3 did not have different configurations. Load was simulated to exert the transaction creation throughput continuously. The code for the TX creating machines and the experiments are
\ifReview{
available\footnote{\href{https://doi.org/10.5281/zenodo.7470507}{https://doi.org/10.5281/zenodo.7470507}}.
}{
available\footnote{\href{https://github.com/OleDe/ethereum-tx-scaling}{https://github.com/OleDe/ethereum-tx-scaling}}.
}

\subsection{Transaction Throughput} \label{sec:txThroughput}

In this section, we discuss the evaluation of the transaction throughput of our approaches to scaling transaction-creating machines. Throughput, in this case, refers to how many transactions are created per second (tps).

\begin{figure}[!b]
    \centering
    \includegraphics[width=.45\textwidth]{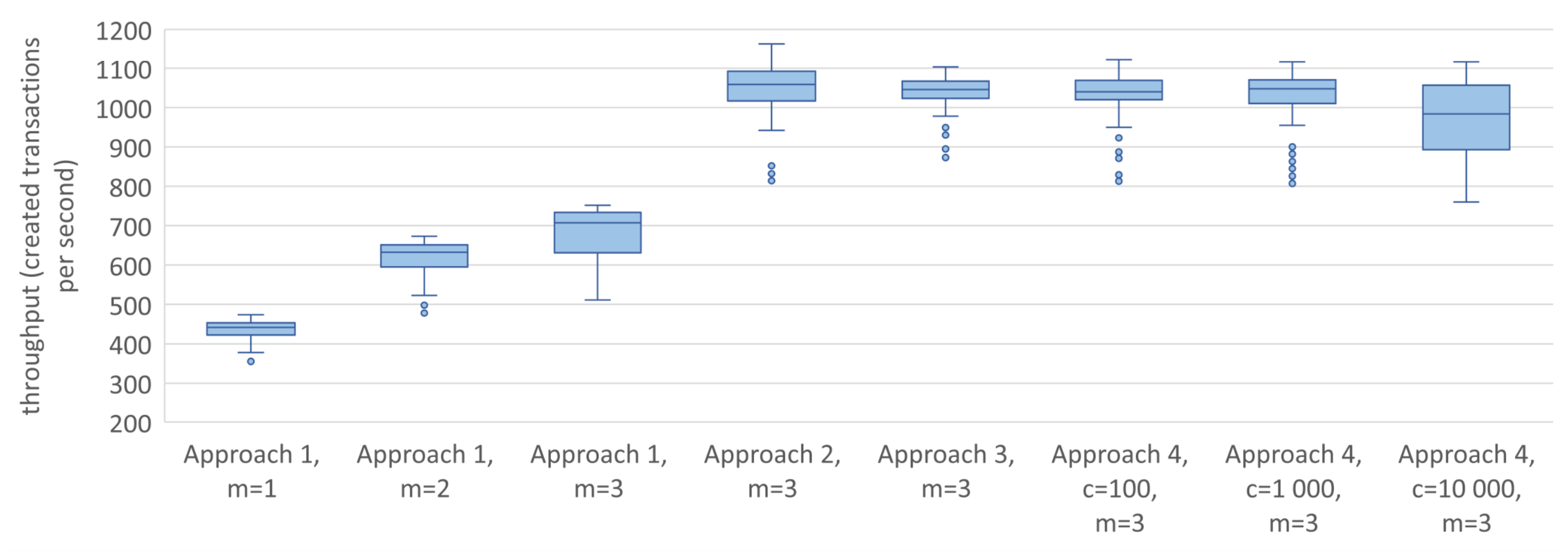}
    \caption{Comparison of the total transaction throughput}
    \label{fig:creation-comparison-total}
\end{figure}

\begin{figure}[!b]
    \centering
    \includegraphics[width=.45\textwidth]{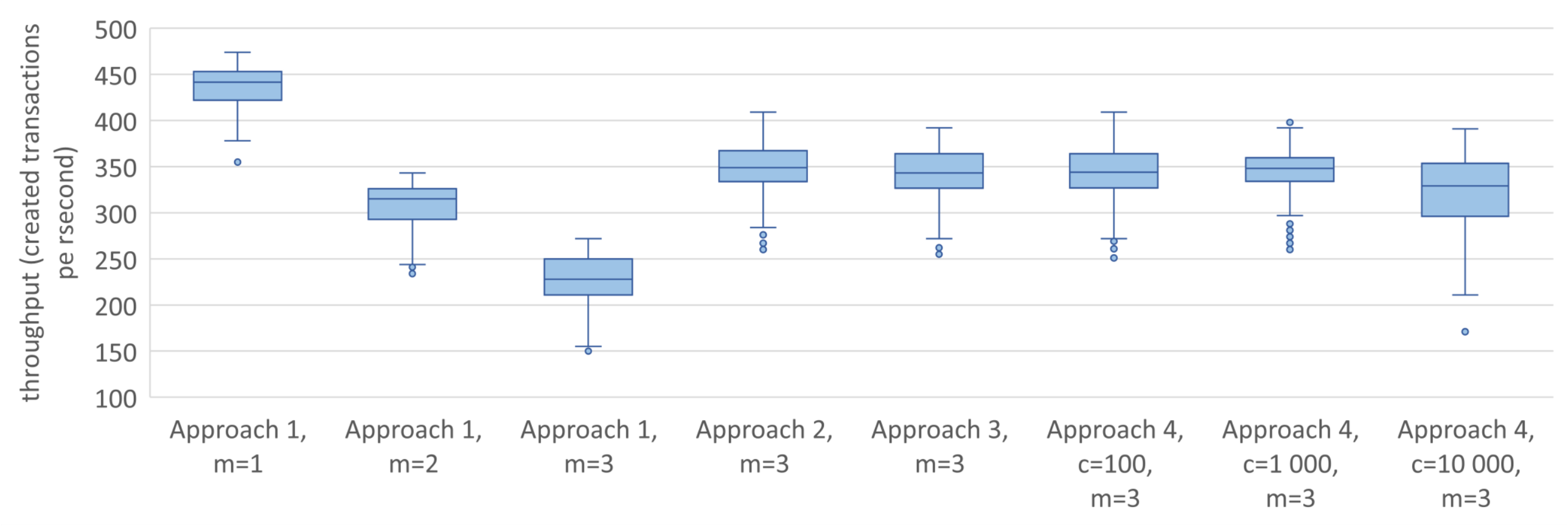}
    \caption{Comparison of the transaction throughput per machine}
    \label{fig:creation-comparison-perMachine}
\end{figure}

Figure \ref{fig:creation-comparison-total} shows a comparison of the total transaction throughput, i.e., the transaction throughput of all machines combined, between the four approaches. We can see that Approach 1 performed the worst. Even with $m=3$ transaction-creating machines running simultaneously, this approach yielded a median of only about $680$ tps. Also, the increase in throughput is sub-linear with a growing number of machines. 
In contrast, all the other approaches achieved significantly higher throughput rates, with medians ranging from $980$ up to $1\,045$~tps. It appears that the middleware used in Approach 1 already poses a severe bottleneck when we only use a small number of machines. Thus, it is not a well-suited solution for horizontally scaling transaction creation. We observed the highest throughput for Approach~2 with $1\,045$~tps (median). That corresponds to our expectations because the transaction-creating machines in Approach 2 can locally keep track of the current nonce. In contrast, Approaches 3 and 4 regularly have to make external requests to the Nonce Manager. Still, the difference in performance between Approaches 2, 3 and 4 is only marginal. It is especially notable that the median transaction throughput for Approach 3 with $1\,030$ tps was only slightly smaller than that of Approach 2, even though the transaction-creating machines had to make a request to the Nonce Manager for every single transaction they created.
Figures~\ref{fig:creation-comparison-total} and~\ref{fig:creation-comparison-perMachine} show a similar spread of the distributions for all settings, with two exceptions: Approach 1, $m=3$ and Approach 4, $c = 10\,000$. The former is clearly suboptimal, and does not warrant further discussion, while the latter is of interest and this setting will be discussed in detail in Sect.~\ref{sec:txWaitingPeriods}.

Figure \ref{fig:creation-comparison-perMachine} shows the transaction throughputs per machine for the different approaches and configurations. Here we can see even more clearly that the middleware in Approach 1 becomes a bottleneck as soon as we try to horizontally scale transaction creation. When only a single transaction-machine was running, Approach 1 achieved a median throughput of $435$ tps, the highest observed throughput across all machines and configurations. The reason for this is that the machines in Approach 1 do not have to sign the transactions, but instead only create and forward ``unfinished'' transactions to the middleware. For all other approaches, the machines are themselves responsible for signing the transactions, which leads to lower performance in the individual machine. However, once we consider $m=2$ transaction-creating machines, the throughput per machine of Approach 1 drops to a median of $307$ tps. At $m=3$ machines, Approach 1 only offers a median throughput per machine of $230$ tps, which is significantly lower compared to the other approaches, whose median throughput rates per machine ranged from $323$ tps to $348$ tps.

Summarized, Approaches 2, 3 and 4 offer adequate throughput rates that increase proportionally when adding more transaction-creation machines. For Approach 1, on the other hand, the middleware poses a bottleneck when scaling horizontally, so its throughput does not increase proportionally when the number of machines is increased. However, for all four approaches, the throughput was mostly stable, i.e., the variance of the throughput over time was low. In addition, the load was distributed evenly across all active machines; no single machine had significantly higher/lower throughput than others during the same run.

\subsection{Latency and Waiting Periods} \label{sec:txWaitingPeriods}

In this section, we examine the transactions' latency, measured as \emph{waiting periods}. With the term \emph{waiting period} we describe the time it takes from a transaction's creation at the transaction-creating machine to it being included in the Ethereum blockchain. Note that we do not use the term \textit{commit time} from the literature~\cite{2017-Weber-SRDS}, since that is measured from transaction announcement to the network, which is less suitable for our purposes.

\begin{figure}[tb]
    \centering
    \includegraphics[width=.45\textwidth]{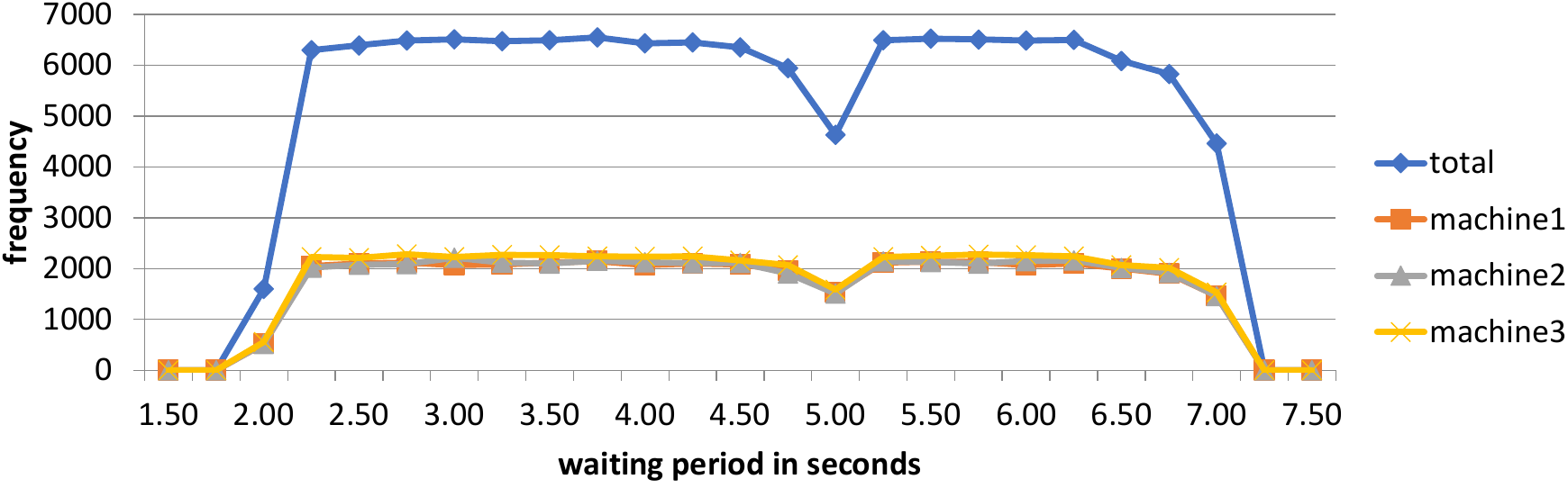}
    \caption{Approach 2: aggregated waiting periods with bin sizes of $0.25$ seconds}
    \label{fig:waitingTimes-histogram-2}
\end{figure}
\ifReview{}{
\begin{figure}[tb]
    \centering
    \includegraphics[width=.45\textwidth]{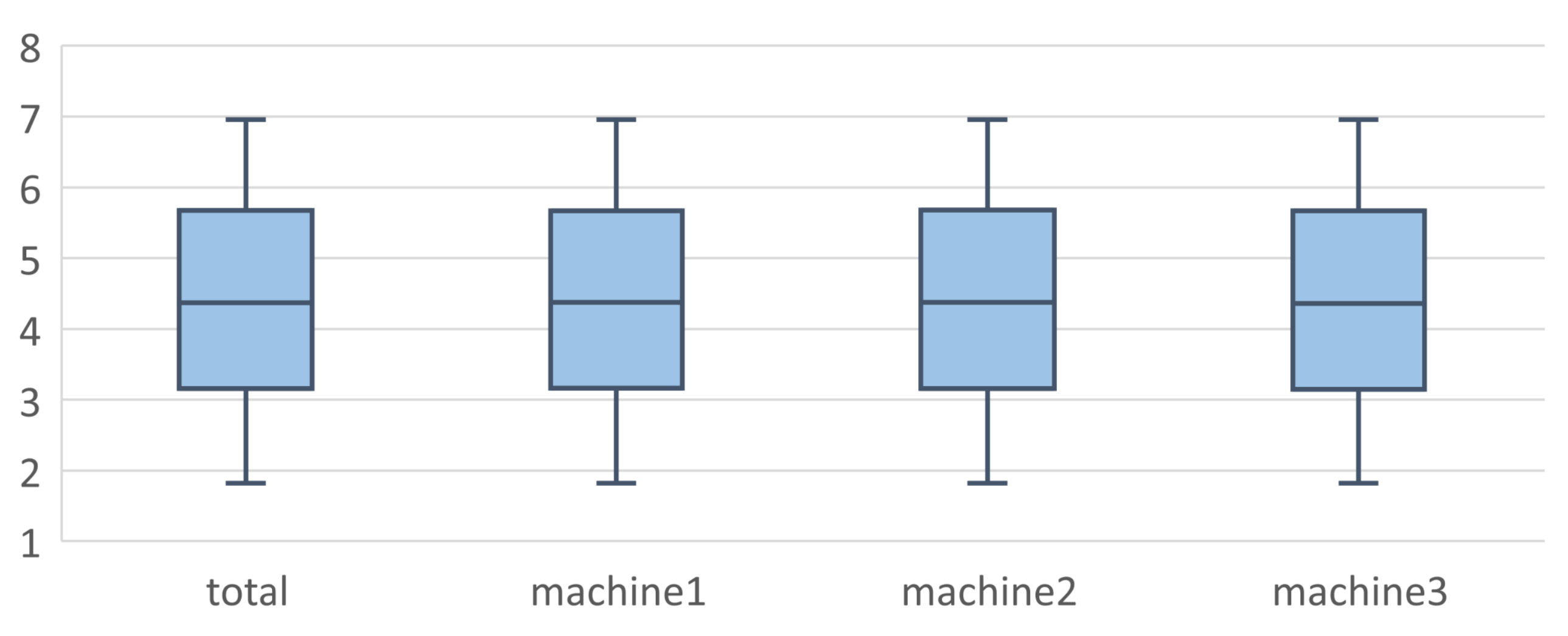}
    \caption{Approach 2: Boxplots of waiting periods (in seconds)}
    \label{fig:waitingTimes-boxplot-2}
\end{figure}
}

\ifReview{Figures \ref{fig:waitingTimes-histogram-2} shows the distribution of waiting periods for Approach 2; observations for Approaches 1 and 3 were almost identical, hence the following also applies to them.}{Figures \ref{fig:waitingTimes-histogram-2} and \ref{fig:waitingTimes-boxplot-2} show the distribution of waiting periods for Approach 2; observations for Approaches 1 and 3 were almost identical, hence the following also applies to them. As can be seen from Figure~\ref{fig:waitingTimes-boxplot-2}, the distributions are very similar; hence we focus on discussing the medians.}


The shortest (longest) observed waiting periods were at approximately $1.8$ ($7$) s, respectively. In-between this interval, the waiting periods follow roughly a uniform distribution. This is to be expected, because we continuously create new transactions while creating new blocks and thereby including transactions in the blockchain every $5$ s. Hence, the length of the interval also approximately corresponds to the configured inter-block time of $5$ s. On average, it took $4.4$ s from a transaction's creation to its inclusion in the blockchain.

The frequency of observed waiting periods noticeably drops at the $5$-second-mark, as can be seen in Fig.~\ref{fig:waitingTimes-histogram-2}. Through further testing we could confirm that this drop depends on the configured inter-block time, i.e., when changing the inter-block time to a specific value, the drop could then be observed around that new inter-block time. 
Given that the behavior of Ethereum clients is not our focus, we did not investigate this particularity further.

\begin{figure}[tb]
    \centering
    \includegraphics[width=.45\textwidth]{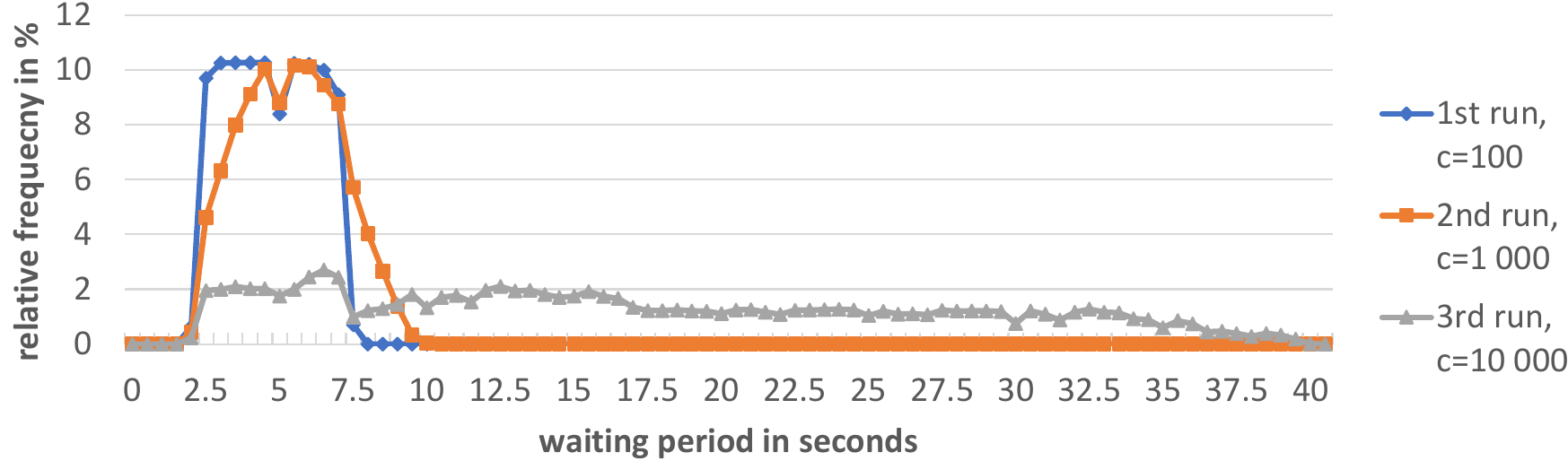}
    \caption{Approach 4: aggregated waiting periods with bin sizes of $0.5$ seconds for different contingent sizes}
    \label{fig:waitingTime-comparison-4}
\end{figure}

Figure \ref{fig:waitingTime-comparison-4} compares the distribution of waiting times of Approach 4 for contingent sizes of $c=100$, $c=1\,000$ and $c=10\,000$. In general, the observed waiting periods were lower when smaller contingent sizes were used. That is expected behavior, because assigning nonce contingents will lead to situations where transactions with higher nonces have already reached the transaction pool of our Ethereum node, while some transactions with smaller nonces have not yet been created. The transactions with the higher nonces then have to wait in the queue until all transactions with lower nonces have been created and sent to the Ethereum Node as well. For larger nonce contingents, this effect intensifies and the average length of the waiting period increases.

For a contingent size of $c=100$, the distribution of waiting periods is similar to that of Approaches 1, 2 and 3. The observed waiting periods range from $1.8$ to $7.2$ s. Within this interval, they approximately follow a uniform distribution, except for the drop in frequency around the $5$-second-mark which we also observed for the other approaches. When increasing the contingent size to $c=1\,000$, the average length of the waiting periods already slightly rises. But for contingent sizes of $c=10\,000$ we observe a drastic upsurge, with transactions waiting up to $40$ s from their creation to being included. With a transaction throughput of about $330$ tps on average, the transaction-creating machines are fast enough to completely deplete a whole contingent for $c=100$ and $c=1\,000$ within the $5$ s inter-block time. But for $c=10\,000$, a machine needs about $30$~s until it has used all nonces of a contingent. Accordingly, for larger contingent sizes, the inter-block time is less and less the determining factor for a transaction's estimated waiting period, but instead the average time it takes for all transactions with lower nonces to also reach the miner.

Nevertheless, as long as the contingent size is not configured to be overly large, Approach 4 offers comparable latency / waiting periods to the other approaches. Across all approaches, waiting periods were low and independent from the transaction-creating machine from which a transaction originated.

\subsection{Performance of the Mining Node}

In this section, we present additional data regarding the inclusion of transactions in the blockchain and the status of our miner in general. Like in Section~\ref{sec:txWaitingPeriods}, we first discuss results for the first three approaches and then regard Approach 4 separately. Again, Approaches 1, 2 and 3 behaved very similarly, hence Figures \ref{fig:txMiningRate-2} and \ref{fig:txPool-2} are representative for Approaches 1 and 3 despite being based on data from Approach 2.

\begin{figure}[tb]
    \centering
    \includegraphics[width=.45\textwidth]{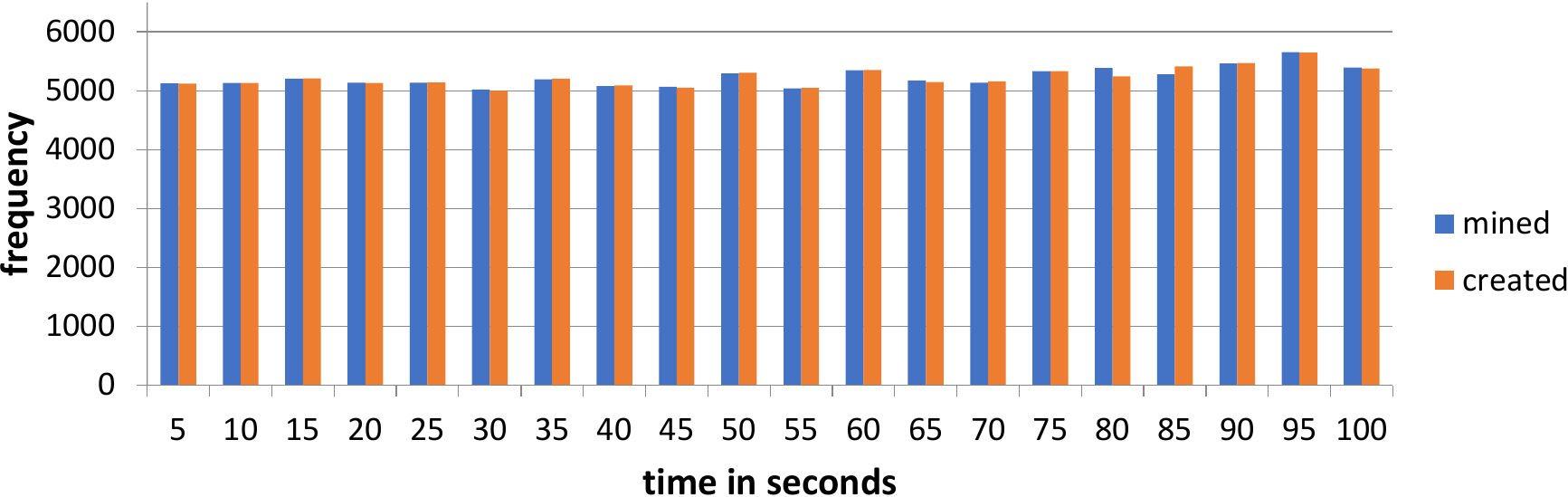}
    \caption{Approach 2: Comparison between the number of created transactions and the number of transactions included in the blockchain over intervals of 5 seconds; for tps, divide by 5.}
    \label{fig:txMiningRate-2}
\end{figure}

\begin{figure}[tb]
    \centering
    \includegraphics[width=.45\textwidth]{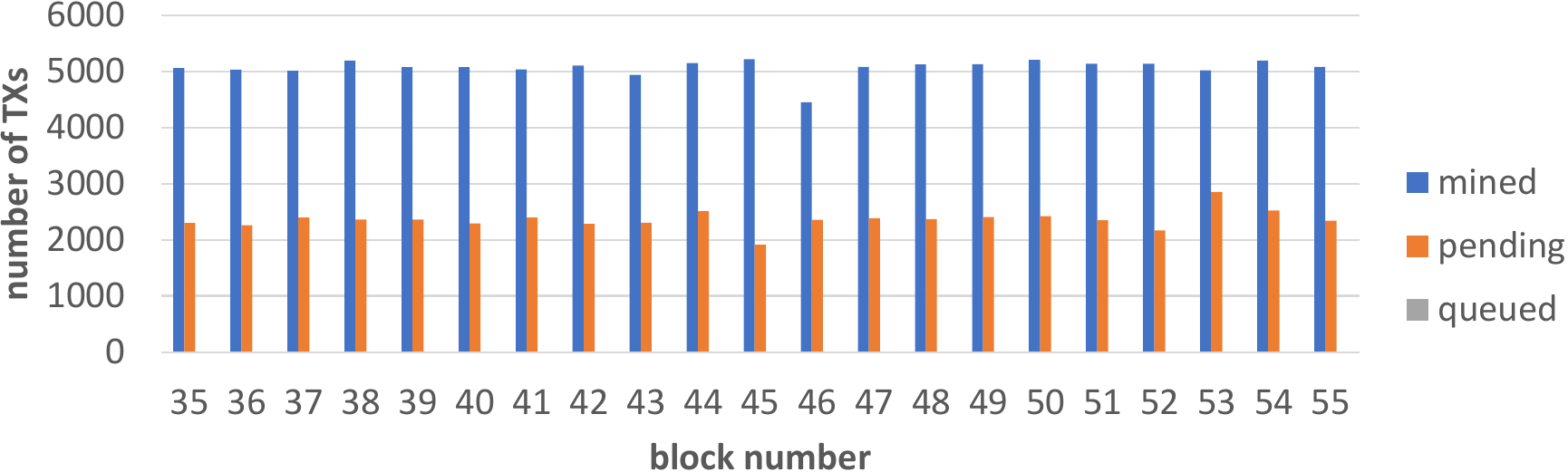}
    \caption{Approach 2: transaction count per block compared to the number of pending and queued transactions right after the block was mined}
    \label{fig:txPool-2}
\end{figure}

Figure \ref{fig:txMiningRate-2} depicts a comparison between the number of created transactions and the number of transactions included in the blockchain over intervals of $5$~s, respectively. We chose $5$ s as the size for the intervals because it corresponds to the inter-block time of $5$ s. We can see that both the mining rate (blue) and the creation rate (orange) are quite stable, with no noteworthy fluctuation. In addition, the mining rate is approximately identical to the creation rate, so the miner was able to constantly keep up with the transactions-creating machines. That is also visible in Figure \ref{fig:txPool-2}, where we can see the transaction count per block compared to the size of the transaction pool (pending transactions and queued transactions) right after the block was mined. The observed number of pending transactions did not increase over time but almost stayed constant, so the transactions were included in the blockchain just as fast as they were created. The number of queued transactions was always zero. This is expected, since, for Approaches 1, 2, and 3, the transactions are always created in sequence according to their nonce. Therefore, we observed no instance of a transaction with a higher nonce being in the transaction pool before a transaction with a lower, at the points of observation.

The same does \emph{not} apply for Approach 4, especially for larger contingent sizes, as can be seen in Figure \ref{fig:txMiningRate-4-10000}. While the transaction creation rate is fairly stable, the contingent size of $c=10\,000$ led to heavy, periodic fluctuations for the rate at which transactions are included in the blockchain. Here, we observed a standard deviation of $\sigma = 3\,800$.

The reason for these fluctuations can be seen in Figure \ref{fig:txPool-4-10000}: when using larger contingent sizes $c$, transactions with higher nonces remain in the transaction pool longer, in the \emph{queued} status. They have reached the Ethereum node but cannot be included (or ``mined'') because not all transactions with lower nonces have yet been created and sent to the Ethereum node. The maximum possible number of these temporarily ``missing'' transactions is of course higher for larger contingent sizes $c$. More specifically, with $m=3$ transaction-creating machines running at roughly the same speed, the maximum is at $2c$. The higher the value of $c$, the longer it takes the transaction-creating machines to create all the missing transactions, and the longer the queued transactions with the higher nonces have to wait and their number is building up. Eventually, when all missing transactions have been created and sent to the Ethereum node, the queued transactions can be included in the blockchain. Therefore, they are shifted to the \emph{pending} status all at once. They are then included in the next block (or in the next multiple blocks, if their combined amount of gas is higher than the gas limit for a single block) to be mined and become a part of the blockchain. As a result, the mining rates are periodically fluctuating above and below the creation rates in Figure \ref{fig:txMiningRate-4-10000}. The theoretical maximum of simultaneously queued transactions is also $2c$, assuming we have $m=3$ transaction-creating machines with the same throughput.

\begin{figure}[t]
    \centering
    \includegraphics[width=.45\textwidth]{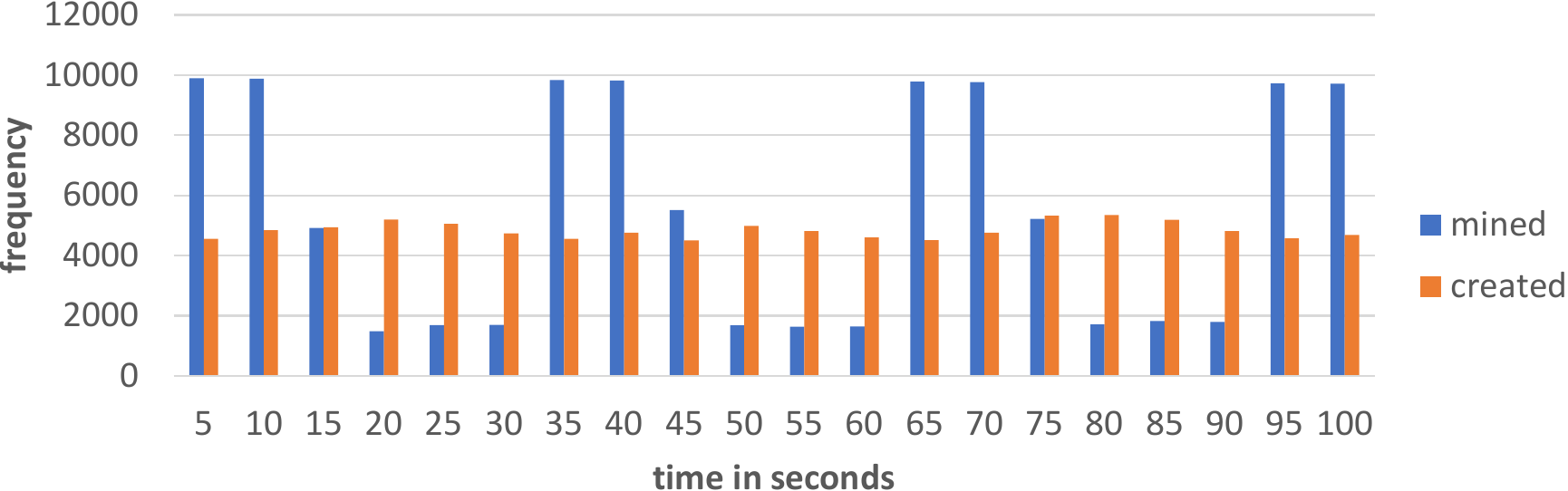}
    \caption{Approach 4, $c=10\,000$: Comparison between the number of created transactions and the number of transactions included in the blockchain over intervals of 5s}
    \label{fig:txMiningRate-4-10000}
\end{figure}

\begin{figure}[t]
    \centering
    \includegraphics[width=.45\textwidth]{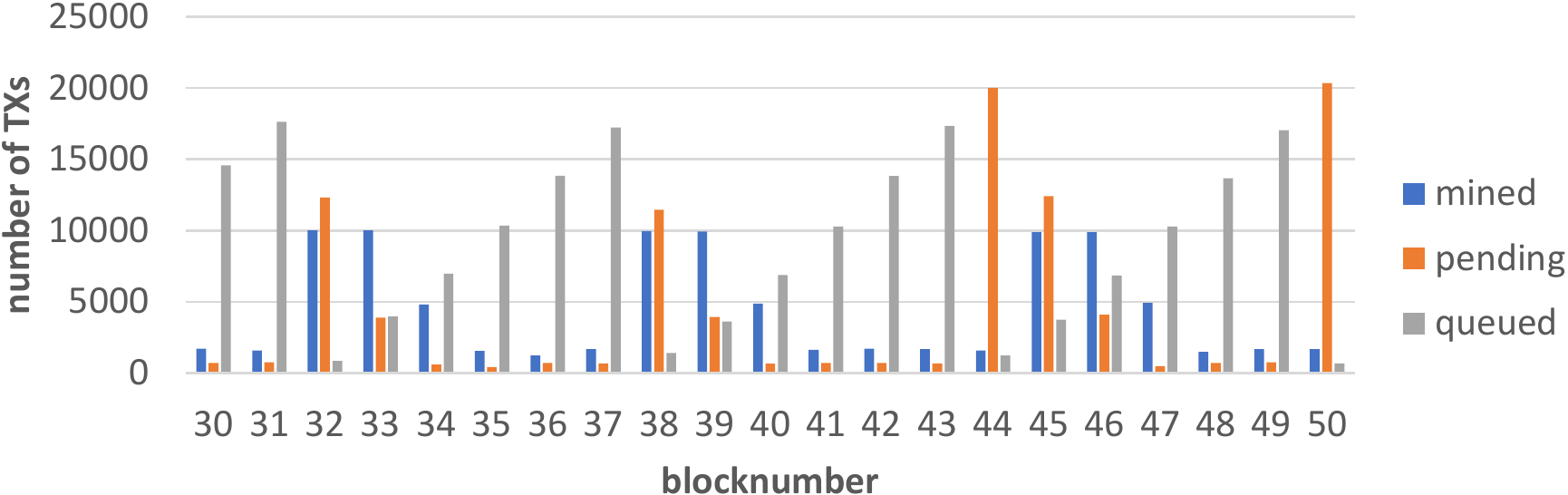}
    \caption{Approach 4, $c=10\,000$: transaction count per block compared to the number of pending and queued transactions right after the block was mined}
    \label{fig:txPool-4-10000}
\end{figure}

\section{Discussion}
\label{sec:discussion}

Overall, the first approach, where we add the nonce at a dedicated singleton \emph{middleware}, displayed the weakest scalability. This was somewhat expected, but determining the points when the scalability limitations of this approach would become relevant (if at all) was an interesting object that deserved investigation.
While the transaction throughput increased proportionally for additional transaction-creating machines in Approaches 2, 3, and 4, this was not the case for Approach 1. Instead, the throughput increase caused by adding additional machines quickly declines, so scaling from two to three machines only increased throughput by $10.8$ \%. By examining Approach 1, we explored its limitations to determine which for use cases (number of VMs, number of tps) it remains suitable.
When scalability is a main objective, Approach 1 should likely be avoided.

In contrast, Approach 2 displayed a much more favorable performance, with a very stable throughput that sufficiently increases when adding transaction-creating machines. The key advantage of Approach 2 is that the transaction-creating machines do not depend on any other component to set the nonce or to forward transactions to the Ethereum Node, which is why it offered the overall highest throughput of $1\,045$ tps. As a downside, Approach 2 requires supplying a distinct Ethereum Account for each transaction-creating machine and a slightly higher gas usage. This might introduce higher complexity for dynamically scaling the number of machines and for realizing authentication and authorization. In summary, Approach 2 is a solid design choice to implement horizontal scalability for transaction-creating machines, provided that using multiple Ethereum accounts or higher gas usage are acceptable in a given context.

For Approach 3, it is harder to give a clear recommendation. During our simulation run with three transaction-creating machines, the transaction throughput of our implementation was similar to Approach 2. With an average of $1\,030$ tps, it even outperformed Approach 4 -- albeit slightly -- for all three tested contingent sizes. However, the singleton Nonce Manager will eventually become a bottleneck when too many transaction-creating machines are running concurrently, so both Approach 3 and Approach 4 cannot be scaled indefinitely. The big disadvantage of Approach 3 is that the transaction-creating machines will have to make $c$-times as many requests to the Nonce Manager as they would in Approach 4. This means the Nonce Manager will become a bottleneck in Approach 3 considerably sooner, so the maximum number of machines running concurrently is drastically higher in Approach 4.  
In contrast to the other approaches, if the contingent size $c$ is too large, Approach 4 is suboptimal in terms of latency, and particularly the distribution of waiting times, which can be understood as \textit{fairness}. By selecting a smaller $c$, this issue can be avoided.
Given these points, Approach 4 should generally be preferred over Approach 3, even though the average throughput was slightly higher for Approach 3.

Approaches 2 and 4 are both decent choices for implementing horizontal scaling of transaction creation, and they offer very similar performance. However, while Approach 2 requires a way to deal with account management and authorization (see Section~\ref{sec:approach2}), Approach 4 necessitates implementing an additional component -- the Nonce Manager -- which poses a single point of failure, so there need to be adequate measures for mitigation and recovery. 
For a given context, the choice comes down to finding the better trade-off between the drawbacks and advantages, in most contexts likely between Approaches 2 and 4.

\section{Conclusion and Future Work}
\label{sec:conclusion}
Following the increased throughput scalability of blockchain systems, high-volume applications with a single participant issuing thousands of transactions per second become possible.
However, the horizontal scaling of transaction-creating machines has received little attention in research to date.
We study this subject and propose four approaches to this end. Prototypical  implementations of the approaches allowed us to evaluate them experimentally.
Our work demonstrates that it is feasible to horizontally scale transaction creation, and two of the approaches -- Approaches 2 and 4 -- achieved both good scalability and fair latency distributions.
These two approaches come with different trade-offs that need to be considered in a real-world implementation. 
Approach 2 introduces additional complexity because it requires on-chain account management and slightly higher gas usage, but it offers the highest throughput. 
Approach 4 relies on the Nonce Manager, which could become a single point of failure, and achieved a throughput only slightly below that of Approach 2.

\ifReview{In future work, we plan to consider the handling of failures and ordering constraints on the incoming requests. In addition, we plan to extend the evaluation in various directions, including resiliency and other configurations of networks. Finally, we plan to investigate additional approaches and further optimizations.}
{There are some additional factors 
that should be addressed in further research. One open point is handling the failure of transactions. Should a single transaction for some reason not be included in the blockchain, then all following transactions cannot be included either.
Thus, the work can be amended with a monitoring approach to detect failed transactions and re-try, e.g., as per the method in~\cite{2017-Weber-SRDS}.
Second, we assumed there are no dependencies between transactions other than the nonce of their associated Ethereum account. 
However, the business logic of Smart Contracts might require the transactions to be processed in a specific sequence, and that sequence would then need to be accounted for when setting the nonces during transaction creation. This case should be researched more, as it might require changes to the transaction creation approaches, or even render Approach 4 unsuitable. Capturing dependencies between transactions, e.g.\ as in~\cite{Reijsbergen2020}, might be a viable starting point.
}

\bibliographystyle{ieeetr}
\bibliography{references,lit}


\end{document}